\documentclass[reprint,superscriptaddress,amsmath,amssymb,aps,prl]{revtex4-1}

\usepackage{graphicx}
\usepackage{dcolumn}% Align table columns on decimal point
\usepackage{bm}% bold math
\usepackage{hyperref}% add hypertext capabilities
\usepackage{amssymb}
\usepackage{amsmath}
\usepackage{color}
\usepackage{txfonts}
\usepackage{bm}
\usepackage{ulem}

\usepackage{color}

\begin{document}

\preprint{APS/123-QED}

\title{Clustering of Topological Charges in a Kagome Classical Spin Liquid}
% Force line breaks with \\

\author{Tomonari Mizoguchi}
\affiliation{Department of Physics, University of Tokyo, Hongo, Bunkyo-ku, Tokyo 113-0033, Japan}
\email{mizoguchi@hosi.phys.s.u-tokyo.ac.jp}
\altaffiliation[Present address: ]{Department of Physics, Gakushuin University, Mejiro, Toshima-ku, Tokyo 171-8588, Japan}

\author{L. D. C. Jaubert}
\affiliation{Okinawa Institute of Science and Technology Graduate University, Onna-son, Okinawa 904-0495, Japan}
\affiliation{CNRS, Univ. Bordeaux, LOMA, UMR 5798, F-33400 Talence, France}

\author{Masafumi Udagawa}
\affiliation{
Department of Physics, Gakushuin University, Mejiro, Toshima-ku, Tokyo 171-8588, Japan
}

\date{\today}

\begin{abstract}
Fractionalization is a ubiquitous phenomenon in topological states of matter. 
In this work, we study the collective behavior of fractionalized topological charges and their instabilities,
through the $J_1$-$J_2$-$J_3$ Ising model on a kagome lattice. This model can be mapped onto a Hamiltonian of interacting topological charges under the constraint of Gauss' law. 
We find that the recombination of topological charges gives rise to a yet unexplored classical spin liquid.
This spin liquid is characterized by an extensive residual entropy, as well as the formation of hexamers of same-sign topological charges. 
The emergence of hexamers is reflected by a half-moon signal in the magnetic structure factor, which provides a signature of this new spin liquid in elastic neutron-scattering experiments. To study this phase, a worm algorithm has been developed which does not require the usual divergence-free condition.

\begin{description}
\item[PACS numbers] 75.10.Kt 
\end{description}
\end{abstract}

\pacs{}

\maketitle
Fractionalization is a hallmark of topological states of matter. 
In these systems, an excitation with a unit quantum number, such as a charge and a spin, 
is fractionalized into several constituents.
These excitations can then condense into exotic topological phases~\cite{kitaev2001}.
The nature of fractionalized excitations have been studied through a number of systems, such as quasi-one-dimensional conducting polymers~\cite{Heeger1988}, fractional quantum Hall systems~\cite{Laughlin1983,Goldman1995},
and quantum spin liquids (QSLs) in one and higher dimensions~\cite{balents2010}.

Among the systems showing fractionalization, QSLs are of special interest.
QSL is a long-range-entangled quantum ground state without spontaneous symmetry breaking; 
the ground state is expressed as a superposition of a macroscopic number of product states.
The realization of QSL has been intensively sought in frustrated magnets, and a number of candidate materials have been explored actively~\cite{hiroi2001,shimizu2003,shores2005,okamoto2007,okamoto2009,singh2010,fak2012,takayama2015}.

Quantum spin ice~\cite{gingras2014,savary2016} 
is one of the most promising systems in this context.
Its classical counterpart, spin ice, is a classical spin liquid (CSL) with macroscopically degenerate ground states.
CSL often serves as a constituting source of QSL, 
thanks to quantum fluctuations inducing a superposition between the degenerate ground states.
This is why the parent CSL phase reflects several important properties of its descendant QSL state.
In the case of quantum spin ice, a fractional excitation, called monopole, can be found in its classical counterpart, in which a flipped spin from the ground state
is fractionalized into two half-unit charges~\cite{castelnovo2008,jaubert2009}.
CSL, due to its simplicity in comparison with QSL, enables us to study rather intractable aspects of fractional excitations. 

In this work, we study the cooperative phenomena of fractionalized topological charges in the kagome CSL~\cite{kano1953,azaria1987,wolf1988,takagi1993} of the $J_1$-$J_2$-$J_3$ Ising model.
The nearest neighbor (NN) interactions ($J_1$) alone lead to a CSL phase, composed of the constrained configurations of topological charges, 
analogous to the fractional monopole excitations in spin ice. 
The further-neighbor interactions ($J_2$, $J_3$) introduce a NN interaction between the topological charges.
The resulting phase diagram supports a variety of (dis-)ordered phases, 
including what is, to the best of our knowledge, a yet unexplored CSL.
This CSL is composed of coexisting hexamer clusters with an extensive residual entropy. 
Its anomalous spatial structure can be detected through a half-moon signal in magnetic structure factor.

%---------%
\begin{figure}[b]
\begin{center}
\includegraphics[bb= 0 0 612 682,width=\linewidth]{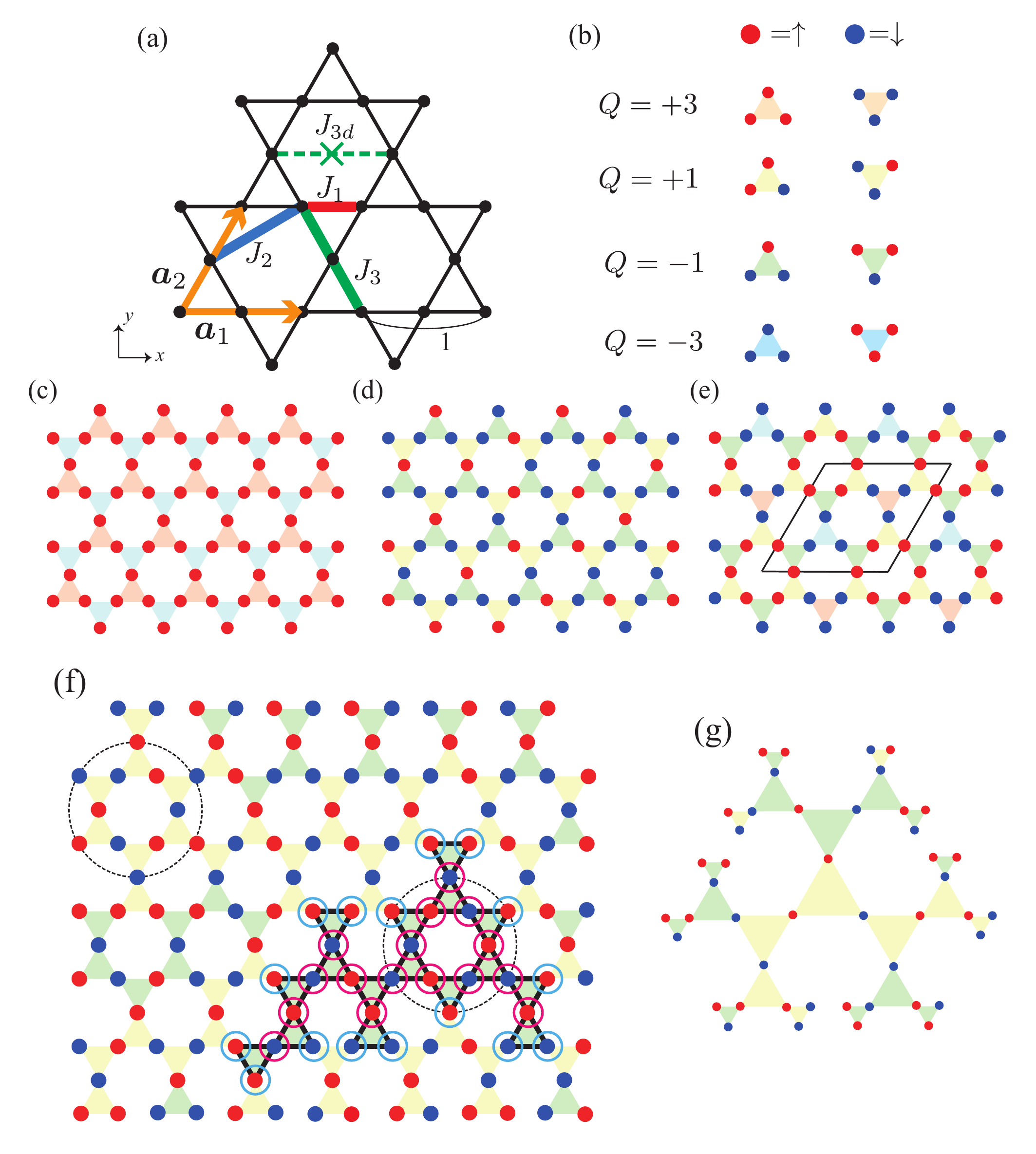}
\vspace{-10pt}
\caption{
(a) Red, blue, and green lines are, respectively, the first-, second-, and third-neighbor interactions on the kagome lattice. Couplings across a hexagon (dashed line) are not considered. The unit vectors $\bm{a}_{i=1,2}$ are represented by orange arrows.
(b) All possible charge states on a triangle with $\sigma^z=+1 (-1)$ marked by red (blue) dots.
(c)-(e)  
The ground state configurations for (c) $J<-1$, (d) $-1<J<0$, (e) $J>\frac{1}{3}$, whose magnetic unit cell is represented by a black rhombus.
(f) One of the ground state configurations for $0<J<\frac{1}{3}$.
The dashed circles denote hexamers.
The region enclosed by the bold line is one of the same-charge clusters,
and pink (light blue) circles denote the internal (boundary) sites of the cluster.
(g) A hexamer CSL state on the Husimi cactus. Even if there are no hexagonal loops, the Husimi cactus allows for each triangle to neighbor exactly two triangles of the same charge, as is the case on average on the kagome lattice. Accordingly, it is a good approximation to compute the thermodynamics of the hexamer CSL [Fig.~\ref{fig2}].}
\label{fig1}
\end{center}
\end{figure}
%---------%

{\it{Model.-}} 
We consider the kagome lattice composed of $N_{\rm site}=L\times L\times 3$ sites under periodic boundary conditions,  
and Ising spin variables, $\sigma_i^z=\pm1$ at each site, $i$.
Throughout this Letter, we set $k_B = 1$.
We define the Hamiltonian as
\begin{align}
\mathcal{H} & = J_1 \sum_{\langle i,j\rangle_{\mathrm{NN} }} \sigma_i^z \sigma_j^z 
+J_2 \sum_{\langle i,j\rangle_{\mathrm{2nd} }} \sigma_i^z \sigma_j^z 
+J_3 \sum_{\langle i,j\rangle_{\mathrm{3rd} }} \sigma_i^z \sigma_j^z,  \label{eq:hamiltonain}
\end{align}
whose couplings are illustrated in Fig. \ref{fig1}(a). 
Here, we assume the NN coupling as antiferromagnetic (AFM), and set its value as a unit of energy, $J_1=1$. We also introduce the second ($J_2$) and the third ($J_3$) neighbor terms. As opposed to a series of recent works on kagome~\cite{Balents2002,Isakov2006,He2014,bieri2015,iqbal2015}, we shall not consider the $J_{3d}$ term across hexagons, but rather the $J_{3}$ term between neighboring triangles which offers, as discussed below, an elegant representation in terms of topological charges.
For general combinations of $J_2$ and $J_3$,
various magnetically ordered states
have been studied~\cite{wolf1988,takagi1993}. 
In the present work, we focus on the case of $J_2=J_3(\equiv J)$, where a charge representation is available~\cite{ishizuka2013_2}.
This model has offered a rich out-of-equilibrium physics on the three-dimensional pyrochlore lattice~\cite{udagawa2016,rau2016}.
To introduce this representation, we classify the triangles on the kagome lattice into ``upward" ($\bigtriangleup$) and ``downward" ($\bigtriangledown$), according to their orientations. 
We then define a topological charge at each triangle as $Q_p = \eta_p \sum_{i\in p} \sigma_i^z $
with $\eta_p=+1(-1)$ for $p\in \bigtriangleup (\bigtriangledown)$  [see Fig. \ref{fig1}(b)], in a manner analogous to the monopole in spin ice~\cite{castelnovo2008}.
This definition naturally leads to the global charge neutrality condition: $\sum_pQ_p=0$.

In terms of $Q_p$, the Hamiltonian can be rewritten as follows, up to a constant term~\cite{ishizuka2013_2}
\begin{equation}
\mathcal{H}  = \left(\frac{1}{2}-J\right)\sum_{p} Q^2_p - J \sum_{\langle p,q \rangle} Q_pQ_q. \label{eq:2}
\end{equation}
The first term represents the self-energy of a charge, summed over all triangles $p$, while the second term is the interaction between neighboring pairs of charges on upward and downward triangles $\langle p,q\rangle$. Opposite charges attract for $J<0$, while they repel for $J>0$. 

{\it{Phase diagram at $T=0$.-}} 
At $J=0$, the model is reduced to the Ising model with AFM NN interaction. Its residual entropy (per site) is $\mathcal{S}_{\mathrm{NN}} =0.502$~\cite{kano1953}.

For finite $J$, this degeneracy is lifted, and 
it becomes instructive to rewrite the Hamiltonian (\ref{eq:2}), as
\begin{equation}
\mathcal{H}  = \frac{1}{2}\left(1+J\right)\sum_{p} Q^2_p - \frac{J}{2}\sum_{\langle p,q \rangle} (Q_p+Q_q)^2. \label{eq:4}
\end{equation}
For $J<0$, 
the energy is minimized by setting (i) $Q_p+Q_q=0$ for all the NN triangle pairs
(i.e., staggered charge orderings),
and (ii) $|Q_p|=1$ ($=3$) on all the triangles for $-1<J<0$ ($J<-1$).

As a result, the ground state is a ferromagnet (FM) with triple-charge ordering for $J<-1$ [Fig. \ref{fig1}(c)], and a single charge ordering for $-1<J<0$  [Fig. \ref{fig1}(d)].
The latter phase shows macroscopic degeneracy with partial spin disorder
i.e., a CSL ground state is realized.
This CSL
is known as kagome ice~\cite{matsuhira2002,udagawa2002,isakov2004_2,aoki2004}
and has been observed in artificial spin ice~\cite{moller2009,chern2011,canals2016,sendetskyi2016} and itinerant systems~\cite{ishizuka2013}.

On the other hand, the situation for $J>0$ is quite unusual, in that charges of the same sign attract each other, in contrast to the electrodynamics we are used to. 
Naively, this suggests a uniform charge order which would take the form of a large-scale charge segregation because of global charge neutrality, $\sum_{p} Q_{p}=0$. However, charge configurations are constrained by the underlying spin structure. The constraint is best understood as a discrete analogue of Gauss' law. The total charge of a given set of triangles $D$ equals the sum of fluxes $\sigma_i^z\;\eta_{p}$ piercing through the boundary $\partial D$;
\begin{equation}
\sum_{p\in D}Q_p = \sum_{i\in\partial D}\sigma_i^z\;\eta_{p_{D}(i)} \leq \sum_{i\in\partial D} 1=n_{b}^{D}, 
\label{eq:3}
\end{equation}
where $p_{D}(i)$ is a triangle of $D$ including site $i$. The left-hand side of the equation is the discrete Gauss' law, obtained from the fact that each site, $i$, belongs to one upward and one downward triangle. Accordingly, the summation of $\eta_{p_{D}(i)} \sigma_i^z$ vanishes for internal sites of $D$, leaving only the boundary contribution $\partial D$. Since $|\eta_{p_{D}(i)} \sigma_i^z|=1$, the total charge inside $D$ is bounded by the number of boundary sites, $n_{b}^{D}$. This prevents large regions of the same charge, and thus macroscopic charge segregation. Under the Gauss' constraint, how can we minimize the ground-state energy by clustering charges of the same sign ?

To answer this question, let us first consider the limit of small $J$ ($0<J\ll1$), where triple charges are excluded. We define a same-charge cluster as a maximal set of connected triangles bearing the same charge [Fig.~\ref{fig1}(f)]. The energy of the $\alpha$th cluster, made of $N_{\triangle}^{(\alpha)}$ triangles, is
\begin{equation}
E_c^{(\alpha)}=\Bigl(\frac{1}{2}-J\Bigr)N_{\triangle}^{(\alpha)}-J n_{i}^{(\alpha)}+\frac{J}{2}n_{b}^{(\alpha)},
\label{eq:5}
\end{equation}
where $n_{i}^{(\alpha)}$ and $n_{b}^{(\alpha)}$ are, respectively, the number of internal and boundary sites of the cluster [Fig.~\ref{fig1}(f)], which satisfy $3N_{\triangle}^{(\alpha)} = 2 n_{i}^{(\alpha)} + n_{b}^{(\alpha)}$.
The term $\frac{J}{2} n_{b}^{(\alpha)}$ expresses the interaction between clusters, and the factor $1/2$ accounts for double counting.
Any spin configuration can be decomposed as a paving of the lattice by same-charge clusters, where every cluster is surrounded by clusters of opposite charges. The total energy is thus, $E_{\rm tot} = \sum_{\alpha}E_c^{(\alpha)}$.
The Gauss' law (\ref{eq:3}) imposes the inequality, $N_{\triangle}^{(\alpha)}\leq n_{b}^{(\alpha)}$, which provides a lower bound to the total energy, $E_{\rm tot}\geq\frac{1}{2}(1-3J)N_{\triangle}$. $N_\triangle$ is the total number of triangles.
This lower bound is reached when each and every
cluster forms a closed ring, with potential sprouting branches [Fig. \ref{fig1}(f)]. For more details, see the Supplemental Material~\cite{supplement}. By construction, more than one closed ring in a same-charge cluster is impossible. This ground state is macroscopically degenerate and on average, each charge neighbors exactly two charges of the same sign.

%---------%
\begin{figure}[t]
\begin{center}
\hspace{-1cm}\includegraphics[bb= 0 0 466 695,width=8.cm]{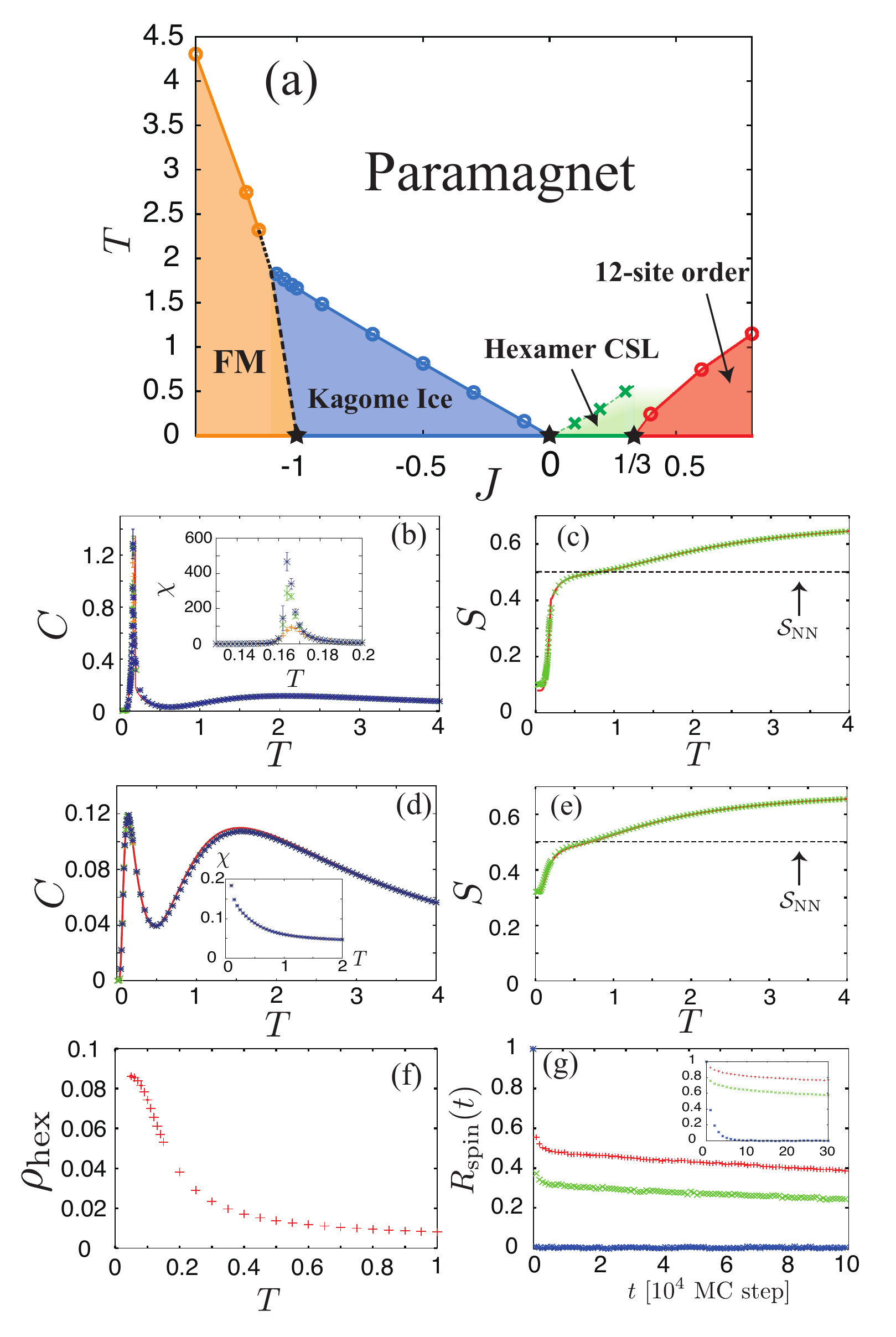}\\
\vspace{-10pt}
\caption{
(a) Finite-temperature phase diagram. Circles and crosses are respectively phase transitions and crossovers. 
Black stars denote the exact phase boundaries at $T=0$. 
(b)-(e) Temperature dependence of (b) $C$ and (c) $S$ at $J=-0.1$, and that of (d) $C$ and (e) $S$ at $J=0.1$.
Insets of (b) and (d) are the magnetic susceptibilities $\chi$.
Red lines are results of Husimi-cactus calculations,
and orange, green, and blue dots are, respectively, the results of Monte Carlo simulations with $L=32$, 64 and $84$.
$S$ is computed for $L=64$.
(f) Temperature dependence of the density of hexamers $\rho_{\mathrm{hex}}$ 
for $J=0.1$ for $L=64$. 
(g) Autocorrelation, $R_{\mathrm{spin}}(t) \equiv \frac{\sum_i \sigma^z_i(0) \sigma_i^z(t)}{N_{\mathrm{site}}}$, for $J=0.1, T=0.03$. 
Red, green, and blue dots are for the single spin flip, the loop-update algorithm, and the new worm algorithm, respectively. }
\label{fig2}
\end{center}
\end{figure}
%---------%

This argument gives rigorous proof that the ground state is a degenerate set of spin configurations where the entire lattice is covered with same-charge clusters containing one and only one ring. As long as this constraint is satisfied, each configuration is realized with equal probability. In principle, rings have no upper-size limit. However, any large ring necessarily encircles other cluster(s), thereby limiting the number of configurations hosting large rings.
Low-temperature classical Monte Carlo simulations confirm this scenario. We find that most of the clusters have hexamer rings, composed of six triangles. We call this novel degenerate ground state the hexamer CSL phase.

It is not possible to pave the entire system with hexamer rings, because this paving resides on a non-bipartite triangular lattice. It means that branches are necessary to accommodate the clusters on the lattice. Then, what is the density $\rho_{\rm hex}$ of hexamer rings defined per number of hexagons in the lattice, $N_{\rm hex}$ ?
Monte Carlo simulations find $\rho_{\rm hex}\sim 0.08$ [Fig.~\ref{fig2}(f)], and thus an 
average number of triangles per cluster, $\frac{N_{\triangle}}{\rho_{\rm hex}N_{\rm hex}}\sim25$. On average, there is an equal number of positive and negative hexamers, but small fluctuations exist.

When increasing $J$, the hexamer CSL shows an instability to triple charge creation. 
By extending the discussion above, one can rigorously show~\cite{supplement} that the phase boundary is at $J=1/3$, with a 12-site order for $J>1/3$ [Fig. \ref{fig1}(e)].

\begin{figure*}[t]
\begin{center}
\includegraphics[bb=0 0 1011 214,width=\linewidth]{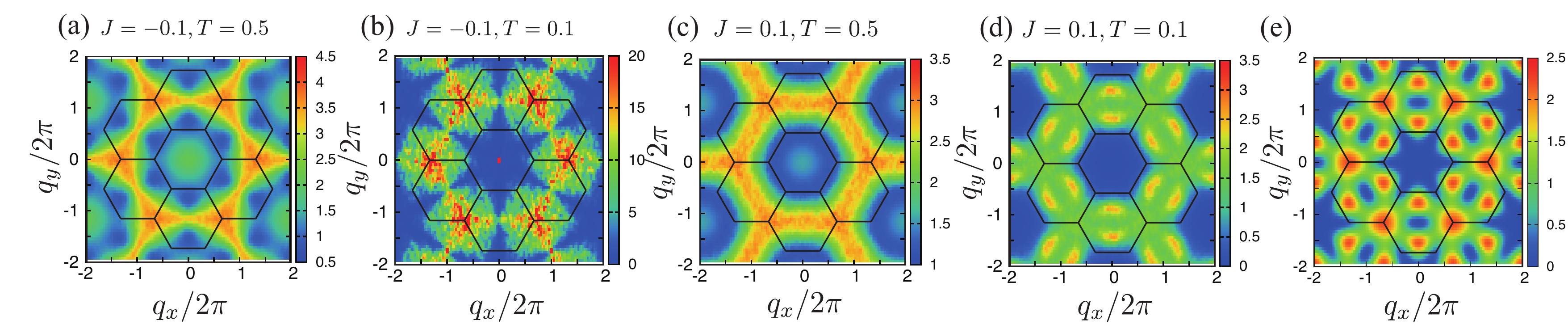}
\vspace{-10pt}
\caption{Magnetic structure factors for $(J,T)=$ (a) $(-0.1,0.5)$, (b) $(-0.1,0.1)$, (c) $(0.1,0.5)$, and 
(d) $(0.1,0.1)$. 
(e) The contribution from isolated hexamers. It confirms that the half-moon patterns observed in (d) are signatures of the hexamer formation. Additional features, such as the scattering at Brillouin-zone corners, are artifacts due to the fact that spins outside of the hexamers are not included in the computation of the structure factor in panel (e).
Black lines denote Brillouin zones. }
\label{fig3}
\end{center}
\end{figure*}
%---------% 
{\it{Thermodynamic quantities.-}} The hexamer CSL for $0<J<1/3$ gives characteristic features in thermodynamic quantities, 
especially when compared to the other disordered regime, kagome ice, for $-1 < J < 0$. 
To see this, we choose $J=-0.1$ and $0.1$, 
and perform a Monte Carlo simulation and Husimi-cactus calculations [Fig.~\ref{fig1}(g)] to obtain the $T$ dependence of the specific heat ($C$), entropy ($S$) and susceptibility ($\chi$) [Figs. \ref{fig2}(b)-\ref{fig2}(e)].
The two methods match quantitatively well. 

$C$ has two peaks both for $J=-0.1$ and $J=0.1$. 
The high-temperature broad peak, around $T\sim1.5$, corresponds to the entropy release due to the vanishing triple charges.
Below this peak, $S$ takes a value close to the residual entropy of the NN Ising model, $S \sim \mathcal{S}_{\rm NN} = 0.502$.

At low temperatures, however, thermodynamic behaviors are quite different between these two regions.
For $J=-0.1$, $C$ and $\chi$ diverge at $T_c \approx 0.163$ due to the long-range charge order with ferrimagnetic spin ordering.
While a previous study interpreted this transition to be of the three-state Potts universality class~\cite{wolf1988}, recent state-of-the-art analyses in the presence of long-range dipolar interactions have shown it to be of the Ising universality class~\cite{moller2009,chern2011}, as expected from the spontaneous $\mathbb{Z}_2$ symmetry breaking of the staggered charge order.

Meanwhile, for $J=0.1$, the specific heat never diverges.
The low-temperature peak is a crossover where the hexamer clusters develop their structure [Fig. \ref{fig2}(f)].
Indeed, $C$ and $\chi$ are almost size independent, indicating the absence of transitions.
The population of hexamer clusters introduces a problem in equilibration, since the cluster is locally quite stable, and a standard
single-spin-flip dynamics freezes. 
Even a loop algorithm~\cite{barkema1998,moller2009,chern2011} effective for the kagome ice region is powerless here, since there is no underlying divergence-free field to support it.
To overcome this difficulty, we developed a new type of worm algorithm~\cite{supplement}, 
which enables nonlocal charge transportation between different clusters, and efficiently equilibrates the system in this region [Fig. \ref{fig2}(g)].

This nonlocal dynamics enables us to explore the thermodynamic behavior below the second peak.
In particular, we succeed in evaluating the $T$ dependence of entropy precisely, 
and find that it converges to an unfamiliar value of $S^{\mathrm{HCSL}}_0 \approx 0.32$, near zero temperature.
This residual entropy is well reproduced by applying a variant of Pauling's argument~\cite{pauling35}
to a nine-spin cluster made of four connected triangles.
As a Pauling constraint, we impose that each triangle neighbors with two, and only two, same-sign charges, giving the entropy
$S_P^{\mathrm{HCSL}} = \frac{1}{6}  \log \left(  \frac{27}{4} \right)\approx 0.32$~\cite{supplement}.

{\it{Magnetic structure factor.-}} 
The unusual spatial structure of the CSLs can be detected by analyzing their magnetic structure factors,
$S(\bm{q}) \equiv \sum_{i,j} \langle \sigma^z_i \sigma^z_j\rangle e^{-i\bm{q}\cdot (\bm{r}_i -\bm{r}_j)}$. 

The spatial structure of kagome ice region is characterized by the singularity of $S(\bm{q})$, called pinch point~\cite{Youngblood81,isakov2004}.
In fact, upon cooling, $S(\bm{q})$ develops bow-tie structures in the centers of the second Brillouin zone, even above $T_c$ [Fig. \ref{fig3}(a)].
Below $T_c$, where the kagome ice correlation is well developed, pinch-point singularities 
are observed [Fig. \ref{fig3}(b)], coexisting with the Bragg peaks of ferrimagnetic order~\cite{canals2016,sendetskyi2016,brooks2014,petit2016,paddison2016}.

In contrast, at $J=0.1$, $S({\bm q})$ develops a quite different pattern. First, the absence of Bragg peaks confirms the absence of long-range order at low temperature.
Pinch points are missing, clearly indicating the hexamer CSL phase is not a Coulomb phase.
Instead, one can see ``half-moon" patterns surrounding the zone centers [Fig. \ref{fig3}(d)].
These patterns give the evidence for the hexamer rings. 
Actually, the contribution from isolated hexamers, $S_{\mathrm{hex}}(\bm{q}) = \sum_{i,j \in \mathrm{hexamers} } \langle \sigma_{i}^z \sigma_j^z \rangle e^{-i \bm{q} \cdot (\bm{r}_i-\bm{r}_j)}$,
hosts a broader version of the half-moon patterns [Fig. \ref{fig3}(e)].
Similar patterns are observed on a fine-tuned point of the $J_{1}$-$J_{2}$-$J_{3}$ model on pyrochlores~\cite{udagawa2016,rau2016}.

{\it{Discussion and Summary}.-}
We have studied the collective behaviors of topological charges and their instabilities, through the $J_1$-$J_2$-$J_3$ Ising model on the kagome lattice.
We found that the fractionalized topological charges are recombined into a novel CSL, as a result of a keen interplay between the topological constraint and interactions. 
This hexamer CSL is characterized by the hexamer clusters, whose configurational pattern leads to an unconventional value of residual entropy. 
This spatial pattern can be detected through the half-moon signal in the magnetic structure factor.

The hexamer CSL has a number of interesting features. 
Among them, the self-organization of multiscale structure is remarkable.
While the starting Hamiltonian is homogeneous, each ground state configuration composed of the
hexamer cluster coverings, shows quite inhomogeneous spatial structures.
In the presence of quantum or stochastic dynamical processes, it is anticipated that this multiscale structure leads to rich dynamical spectra, possibly reminiscent of glassy systems. The hexamer CSL also offers a promising \textit{extended} region ($0<J<1/3$) for the search of unconventional QSLs~\cite{Balents2002,Isakov2006,He2014,bieri2015,iqbal2015,bauer14,essafi2016,buessen2016} and quantum order-by-disorder mechanisms~\cite{Chernyshev14a,gotze15a,Chernyshev15,Oitmaa16a} beyond the traditional Heisenberg antiferromagnet. Quantum mechanical perturbations will indeed, to a first approximation, lead to a superposition of 
degenerate configurations of the hexamer CSL.

It is tempting to speculate if the interaction between topological charges underlies hexagonal or small-cluster formation found in a broad range of frustrated magnets. For instance, cluster excitations of antiferromagnetic hexagonal loops, which are also referred to as ``spin molecules", were reported in the three-dimensional spinels $A$Cr$_2$O$_4$ ($A=$Zn, Hg, Mg)~\cite{lee2002,tomiyasu2008,tomiyasu2011} and ZnFe$_2$O$_4$~\cite{kamazawa2003,tomiyasu2011_2} by inelastic neutron scattering. Although the spins are not quite Ising in these materials because of a weak single-ion anisotropy, it would be fascinating for future work to clarify the relation between spin molecules and clustering of topological charges. Indeed, the charge representation is useful for a broad range of spin models on frustrated lattices.
We hope the viewpoint of topological charges and their recombination will be useful for further explorations of novel states of matter in
frustrated magnets, and especially their topological properties.

This work is supported by JSPS KAKENHI (No. 26400339, No. 24340076, No. 15H05852, and No. 15K13533), the Okinawa Institute of Science and Technology Graduate University, and MEXT-Supported Program for the Strategic Research Foundation at Private Universities. T. M. is supported by Advanced Leading Graduate Course for Photon Science (ALPS).
 
%%%%%%%%%%%%%%%%%%%%%

\end{document}

\title{Supplemental Material for \lq \lq Clustering of Topological Charges in a Kagome Classical Spin Liquid"}
\author{Tomonari Mizoguchi}
\affiliation{Department of Physics, University of Tokyo, Hongo, Bunkyo-ku, Tokyo 113-0033, Japan}
 \email{mizoguchi@hosi.phys.s.u-tokyo.ac.jp}

\author{L. D. C. Jaubert}
\affiliation{Okinawa Institute of Science and Technology Graduate University, Onna-son, Okinawa 904-0495, Japan}
\affiliation{CNRS, Univ. Bordeaux, LOMA, UMR 5798, F-33400 Talence, France}

\author{Masafumi Udagawa}
\affiliation{
Department of Physics, Gakushuin University, Mejiro, Toshima-ku, Tokyo 171-8588, Japan
}

\date{\today}% It is always \today, today,
 \maketitle

\section{Determination of the phase boundary in $J>0$}
In this section, we present detailed discussions on the ground-state spin configurations and phase boundaries for $J>0$.
In the main text, we briefly discuss the ground-state structure for $0<J\ll1$, by estimating the energy of same-charge clusters [Eq. (5) in the main text].
Here, we give the details of the derivation, and extend the
 argument to larger $J$, where
triple charges 
are also allowed to exist.
 In Sec. \ref{sec:basic_equations} and Sec. \ref{sec:single_charge_cluster}, we summarize the basic equations and the analysis of ground-state energy for single-charge cluster,
 in a slightly different notation from the main text, for later convenience. In Sec. \ref{sec:triple_charge_cluster}, we describe the derivation of ground-state spin configuration and phase boundary for general $J (> 0)$.

\subsection{Basic equations}
\label{sec:basic_equations}
We base our arguments on the Gauss' law as introduced in Eq. (3) in the main text.
This equation results from the definition of charge defined on a triangle, $p$:
\begin{eqnarray}
Q_p = \eta_p\sum_{i\in p}\sigma_i^z,
\label{eq:charge_definition}
\end{eqnarray}
where $\eta_p=+1(-1)$ for $p\in\bigtriangleup(\bigtriangledown)$. If we consider a set of triangles, $D$, and sum Eq.~(\ref{eq:charge_definition}) over $D$, we obtain the Gauss' law:
\begin{eqnarray}
\sum_{p\in D}Q_p = \sum_{i\in \partial D}\eta_{p_{D}(i)}\sigma_i^z.
\label{eq:Gauss}
\end{eqnarray}
As noted in the main text, $\partial D$ means the sites on its boundary, and $p_{D(i)}$  is
a triangle in $D$ that includes the site $i$.
Here, we have used the fact that each site, $i$, belongs to one upward and one downward triangles. Accordingly, the summation of $\eta_{p_{D}(i)} \sigma_i^z$ vanishes for internal sites of $D$, leaving only the boundary contribution.

If $D$ is composed of $N_{\bigtriangleup}^{D}$ triangles (of both orientations), the number of internal sites, $n_i^{D}$, and the boundary sites, $n_b^{D}$ satisfy the relation:
\begin{eqnarray}
3N_{\bigtriangleup}^{D} = 2n_i ^{D}+ n_b^{D}.
\label{eq:cluster_relation}
\end{eqnarray}
The number of boundary sites, $n_b$, limits the amount of charge in $D$. By taking the absolute value of Eq.~(\ref{eq:Gauss}), we have
\begin{eqnarray}
|\sum_{p\in D}Q_p| \leq \sum_{i\in \partial D}|\eta_{p_{D}(i)}||\sigma_i^z| = \sum_{i\in \partial D}1 = n_b^{D}.
\label{eq:charge_inequality}
\end{eqnarray}
On the basis of Eqs.~(\ref{eq:Gauss}), (\ref{eq:cluster_relation}) and (\ref{eq:charge_inequality}), we can obtain various information regarding the ground states. 

\subsection{Single-charge cluster}
\label{sec:single_charge_cluster}
In this subsection, we summarize the analysis of region: $0<J\ll1$, where only the single charges are allowed to exist.
Let us define a same-charge cluster as a maximal set of connected same-charge triangles.
We take the $\alpha$-th cluster composed of $N_{\bigtriangleup}^{(\alpha)}$ triangles, with $n_i^{(\alpha)}$ internal and $n_b^{(\alpha)}$ boundary sites.
If we apply Eq.~(\ref{eq:charge_inequality}) to this cluster, we have
\begin{eqnarray}
N_{\bigtriangleup}^{(\alpha)}\leq n_b^{(\alpha)},
\label{eq:charge_inequality_singlecharge}
\end{eqnarray}
since we only consider single charges. 

Next, we compute the energy of the $\alpha$-th cluster,
\begin{equation}
E_c^{(\alpha)}=\Bigl(\frac{1}{2}-J\Bigr)N_{\triangle}^{(\alpha)}-Jn_{i}^{(\alpha)}+\frac{J}{2}n_{b}^{(\alpha)},
\label{eq:cluster_energy_singlecharge}
\end{equation}
as in Eq.~(5) in the main text.
This definition allows us to express the total energy of the system, $E_{\rm tot}$, as the summation of $E_c^{(\alpha)}$ over the clusters: $E_{\rm tot} = \sum_{\alpha}E_c^{(\alpha)}$.

By combining Eqs.~(\ref{eq:cluster_relation}), (\ref{eq:charge_inequality_singlecharge}) and (\ref{eq:cluster_energy_singlecharge}), we obtain
\begin{eqnarray}
E_c^{(\alpha)}\geq\frac{1}{2}(1-3J)N_{\triangle}^{(\alpha)}.
\end{eqnarray}
By summing up all the cluster contributions, we obtain the lower bound of the total energy:
\begin{eqnarray}
E_{\rm tot}\geq\frac{1}{2}(1-3J)N_{\triangle}\equiv E^s_{\rm tot},
\label{eq:energy_smallJ}
\end{eqnarray}
with the total number of triangles, $N_{\triangle}$. 
$E^s_{\rm tot}$ gives the lowest energy of the system, given that only single charges exist.
$E^s_{\rm tot}$, in fact, gives the ground state energy for small $J$.

\subsection{Triple-charge cluster}
\label{sec:triple_charge_cluster}
\subsubsection{Structure of triple-charge cluster}
Next, we consider a general value of $J (>0)$, where triple charges are also allowed to exist.
Firstly, we prove that there exists at most one triple charge in a same-charge cluster.
To show this, let us assume there are more than two triple charges in a given cluster.
Then, as a subset of cluster, there must be a string of same-charge triangles with two triple charges on both ends, which we call $\mathcal{D}$.
This set of triangles satisfies $n_b^{\mathcal{D}}=N_{\triangle}^{ \mathcal{D}}+2$. 
Accordingly, we obtain
\begin{eqnarray}
|\sum_{p\in \mathcal{D} }Q_p| = N_{\triangle}^{ \mathcal{D} } + 4 > n_b^ \mathcal{D}.
\end{eqnarray} 
This contradicts Eq.~(\ref{eq:charge_inequality}), and consequently, the hypothesis of more than two triple charges in a cluster is proven to be wrong.
Accordingly, each cluster contains at most one triple charge. 
Hereafter, we call a cluster without triple charge, a s-cluster (sc), and a cluster with one triple charge, a t-cluster (tc).
The $\alpha$-th s-cluster composed of $N_{\bigtriangleup}^{(\alpha)}$ triangles, with $n_i^{(\alpha)}$ internal and $n_b^{(\alpha)}$ boundary sites, satisfies the relation, Eq.~(\ref{eq:charge_inequality_singlecharge}). 
Meanwhile, the $\beta$-th t-cluster composed of $N_{\bigtriangleup}^{(\beta)}$ triangles, with $n_i^{(\beta)}$ internal and $n_b^{(\beta)}$ boundary sites, satisfies
\begin{eqnarray}
N_{\bigtriangleup}^{(\beta)} + 2\leq n_b^{(\beta)},
\label{eq:charge_inequality_triplecharge}
\end{eqnarray}
as a result of applying Eq.~(\ref{eq:Gauss}).

\subsubsection{Cluster energy}
Next, we define a cluster energy in the presence of triple charge, so that the total energy is expressed as the sum of cluster energies:
\begin{eqnarray}
E_{\rm tot} = \sum_{\alpha\in{\rm sc}}E_c^{s,(\alpha)} +  \sum_{\beta\in{\rm tc}}E_c^{t,(\beta)}.
\label{eq:total_energy}
\end{eqnarray}
Here, $E_c^{s,(\alpha)}$ ($E_c^{t,(\beta)}$) is the contribution from a cluster without (with) triple charge. 
In the presence of surrounding triple charges, $E_c^{s,(\alpha)}$ is modified from the formula~(\ref{eq:cluster_energy_singlecharge}).
Even though the cluster does not contain any triple charge itself, it can still interact with neighboring triple charges.
\begin{eqnarray}
\scalebox{0.85}{$\displaystyle
E^{s,(\alpha)}_c = \Bigl(\frac{1}{2}-J\Bigr)N_{\triangle}^{(\alpha)}-Jn_{i}^{(\alpha)}+\frac{J}{2}n_{b}^{ss(\alpha)} + \frac{3J}{2}n_{b}^{st(\alpha)}, $}
\label{eq:single_cluster_energy}
\end{eqnarray}
where $n_{b}^{ss(\alpha)}$ ($n_{b}^{st(\alpha)}$) means the number of boundary sites between single charges (single and triple charges).
They satisfy 
\begin{eqnarray}
n_{b}^{ss(\alpha)} + n_{b}^{st(\alpha)} = n_{b}^{(\alpha)}.
\end{eqnarray}
Accordingly, Eq.~(\ref{eq:single_cluster_energy}) can be rewritten as
\begin{eqnarray}
\scalebox{0.95}{$\displaystyle
E^{s,(\alpha)}_c = \Bigl(\frac{1}{2}-J\Bigr)N_{\triangle}^{(\alpha)}-Jn_{i}^{(\alpha)}+\frac{J}{2}n_{b}^{(\alpha)} + Jn_{b}^{st(\alpha)}. 
$}
\label{eq:single_cluster_energy2}
\end{eqnarray}

A cluster with a triple charge composed of $N_{\triangle}^{(\beta)}$ triangles have $N_{\triangle}^{(\beta)} - 1$ single, and one triple charges.
We assume the triple charge neighbors with $n^{(\beta)}(=0, 1, 2, 3)$ single charges within the cluster, and $3-n^{(\beta)}$ charges of other clusters.
Accordingly, we write the cluster energy as
\begin{align}
&E^{t,(\beta)}_c = \Bigl(\frac{1}{2}-J\Bigr)(3^2 + (N_{\triangle}^{(\beta)}-1)) - J[3n^{(\beta)} + (n_{\rm i}^{(\beta)}-n^{(\beta)})] \nonumber\\
&+ \frac{J}{2}n_{b}^{ss(\beta)} + \frac{3J}{2}n_{b}^{st(\beta)} + \frac{3J}{2}n_{b}^{ts(\beta)} + \frac{9J}{2}n_{b}^{tt(\beta)}. 
\label{eq:triple_cluster_energy}
\end{align}
Here, the first and the second terms come from the self-energy of charges and the interaction within the cluster, respectively.
The other terms denote four possible types of inter-cluster interactions.
$n_{b}^{ss(\beta)}$ and $n_{b}^{tt(\beta)}$ denote the number of sites between single charges and triple charges, respectively;
$n_{b}^{st(\beta)}$ ($n_{b}^{ts(\beta)}$) denotes the number of sites between single (triple) charges in the $\beta$-th cluster and triple (single) charges in other clusters.
These numbers satisfy
\begin{eqnarray}
n_{b}^{ss(\beta)} + n_{b}^{st(\beta)} = n_{b}^{(\beta)} - (3 - n^{(\beta)}),
\end{eqnarray}
and 
\begin{eqnarray}
n_{b}^{ts(\beta)} + n_{b}^{tt(\beta)} = (3 - n^{(\beta)}).
\end{eqnarray}
Accordingly, Eq.~(\ref{eq:triple_cluster_energy}) can be rewritten as
\begin{align}
&E^{t,(\beta)}_c = \Bigl(\frac{1}{2}-J\Bigr)(3^2 + (N_{\triangle}^{(\beta)}-1)) - J[3n^{(\beta)} + (n_{\rm i}^{(\beta)}-n^{(\beta)})] \nonumber\\
&+ \frac{J}{2}[n_b^{(\beta)} - (3 - n^{(\beta)})] + \frac{3J}{2}(3 - n^{(\beta)}) + Jn_{b}^{st(\beta)} + 3Jn_{b}^{tt(\beta)}. 
\label{eq:triple_cluster_energy2}
\end{align}
With these definitions of $E^{s,(\alpha)}_c$ and $E^{t,(\beta)}_c$, 
the total energy can be written as in Eq.~(\ref{eq:total_energy}).

Let us evaluate the total energy. 
By combining eqs.~(\ref{eq:total_energy}), (\ref{eq:single_cluster_energy2}) and (\ref{eq:triple_cluster_energy2}), we have
\begin{align}
&E_{\rm tot} = \sum_{\alpha\in{\rm sc}} \Bigl(\frac{1}{2}-J\Bigr)N_{\triangle}^{(\alpha)}-Jn_{i}^{(\alpha)}+\frac{J}{2}n_{b}^{(\alpha)}\nonumber\\
&+ \sum_{\beta\in{\rm tc}} \Bigl(\frac{1}{2}-J\Bigr)(3^2 + (N_{\triangle}^{(\beta)}-1)) - J[3n^{(\beta)} + (n_{\rm i}^{(\beta)}-n^{(\beta)})] \nonumber\\
&+ \frac{J}{2}[n_b^{(\beta)} - (3 - n^{(\beta)})] + \frac{5J}{2}(3 - n^{(\beta)}) + 2Jn_{b}^{tt(\beta)}.
\label{eq:exact_total_energy}
\end{align}
Here, we have used
\begin{eqnarray}
\scalebox{0.9}{$\displaystyle
\sum_{\alpha\in{\rm sc}} n_{b}^{st(\alpha)} + \sum_{\beta\in{\rm tc}} (n_{b}^{st(\beta)} + n_{b}^{tt(\beta)}) = \sum_{\beta\in{\rm tc}}(3 - n^{(\beta)}).
$}
\end{eqnarray}

Now, we apply the relation (\ref{eq:cluster_relation}), and the inequalities due to Gauss' constraint, (\ref{eq:charge_inequality_singlecharge}) and (\ref{eq:charge_inequality_triplecharge}) to Eq.~(\ref{eq:exact_total_energy}), and obtain the inequality:
\begin{eqnarray}
E_{\rm tot}\geq\frac{1}{2}(1-3J)N_{\triangle} + 4\sum_{\beta\in{\rm tc}}(1-Jn^{(\beta)}).
\end{eqnarray}
The equality holds if (i) $N^{(\alpha)}_{\triangle}=n_b^{(\alpha)}$, (ii) $N^{(\beta)}_{\triangle}+2=n_b^{(\beta)}$ for $\alpha\in{\rm sc}$ and $\beta\in{\rm tc}$,
and (iii) there are no triple charges next to each other, $n_{b}^{tt(\beta)}=0$. Since $E^s_{\rm tot}=\frac{1}{2}(1-3J)N_{\triangle}$ gives the lowest energy given only s-clusters, as defined in Eq.~(\ref{eq:energy_smallJ}), the correction of the total energy due to t-clusters can be evaluated as
\begin{eqnarray}
E_{\rm tot} - E^s_{\rm tot}\geq 4\sum_{\beta\in{\rm tc}}(1-Jn^{(\beta)}).
\label{eq:total_energy_correction}
\end{eqnarray}

\subsubsection{Ground-state energy}
Finally, we discuss the ground-state energy of the system, on the basis of Eq.~(\ref{eq:total_energy_correction}).
Since $n^{(\beta)}\leq3$, Eq.~(\ref{eq:total_energy_correction}) implies that triple charges appear for $J>1/3$.
This is indeed the case; 
the hexamer classical spin liquid (CSL) is destabilized at $J=1/3$, and 12-site cluster ordering
with triple charges is stabilized for $J>1/3$, as we show below.

To prove this, we firstly estimate the lower bound of the righthand side of Eq.~(\ref{eq:total_energy_correction}).
Then, we look for the spin configuration that gives this minimum energy. To this aim, we introduce $M_n$, the number of t-clusters with $n^{(\beta)}=n$.
Since a t-cluster with $n^{(\beta)}=n$ is composed of at least $n+1$ triangles, $M_n$ satisfies
\begin{eqnarray}
M_{\rm tot}\equiv\sum_{n=0}^3(1+n)M_n\leq N_{\triangle}.
\end{eqnarray}
Then, we can rewrite Eq.~(\ref{eq:total_energy_correction}) as
\begin{align}
E_{\rm tot} - E^s_{\rm tot}&\geq 4\sum_{n=0}^3(1-Jn)M_n\nonumber\\
&=-4JM_{\rm tot} + 4(1+J)\sum_{n=0}^3M_n.
\end{align}
For any given value of $M_{\rm tot}$, the total number of clusters $\sum_{n=0}^3M_n$ is minimized, 
if $M_0=M_1=M_2=0$ and $M_3=M_{\rm tot}/4$, leading to
\begin{align}
E_{\rm tot} - E^s_{\rm tot}&\geq(1-3J)M_{\rm tot}\nonumber\\
&\geq
\left\{\begin{array}{ll}
0 & (0<J\leq1/3),\\
(1-3J)N_{\triangle} & (J>1/3).\\
\end{array}\right.
\label{eq:minNRJ}
\end{align}
This inequality shows that the hexamer CSL is stabilized for $0<J\leq1/3$. Meanwhile, it implies a more stable state for $J>1/3$.
Indeed, if one can realize a spin configuration that satisfies above conditions (i), (ii) and (iii), and takes $(M_0,M_1,M_2, M_3)= (0,0,0,N_{\triangle}/4)$, 
it will be the ground state in this region, with the ground-state energy, $E_{\rm tot} - E^s_{\rm tot}=(1-3J)N_{\triangle}$.

%---------%
\begin{figure}[!htb]
\begin{center}
\includegraphics[bb = 0 0 421 206,width=\linewidth]{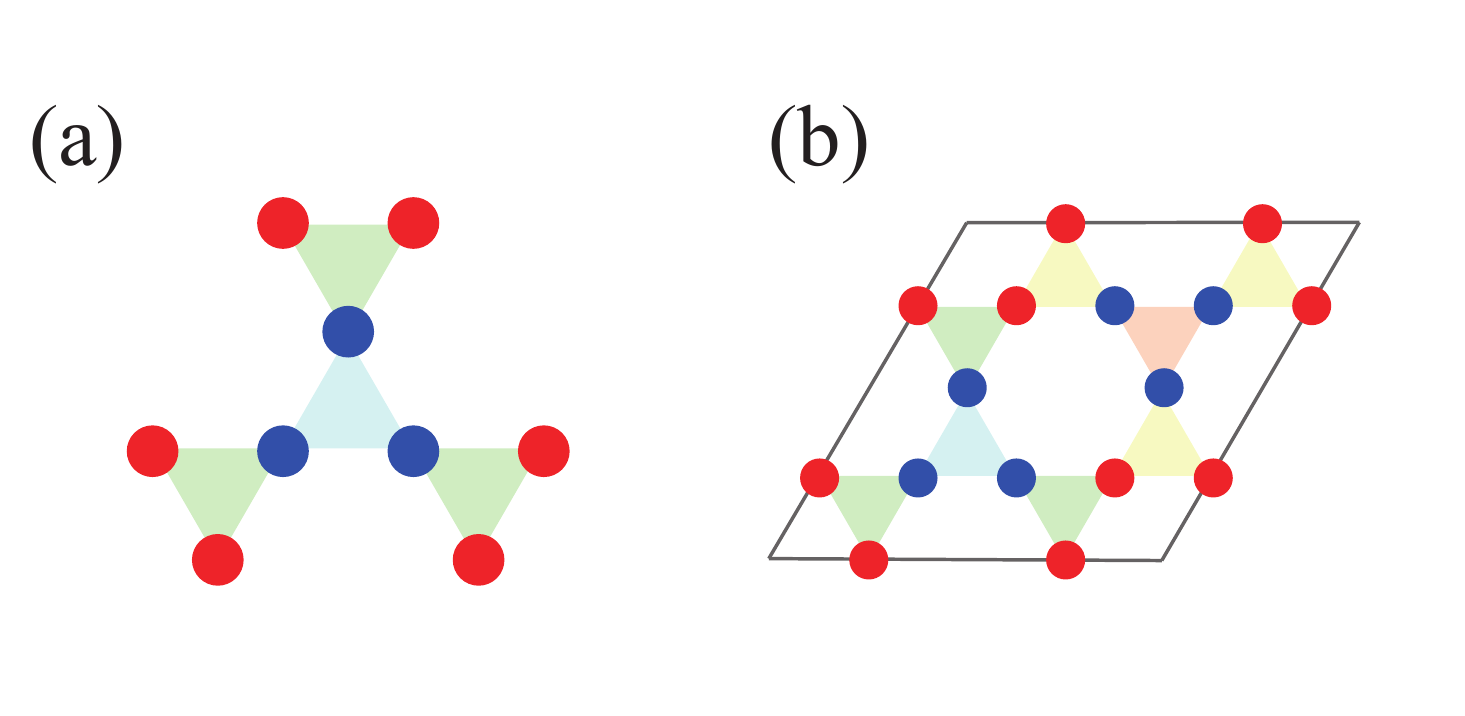}
\vspace{-10pt}
\caption{(a) 4-triangle cluster and (b) a magnetic unit cell of the 12-site ordering. } 
\label{fig:12so}
\end{center}
\end{figure}
%---------%

One can actually realize such spin configuration by packing four-triangle clusters made of a central triple charge, surrounded by three single charges.
[see Fig. \ref{fig:12so}(a)].
To pack these clusters under the constraint of total charge conservation, 
the magnetic unit cell has $2 \times 2$ structure 
with 12 distinct sites, 
thus we call this state \lq \lq 12-site ordering" [see Fig. \ref{fig:12so}(b)].
The energy of this state is given by 
\begin{align}
E_{\rm tot}-E^{s}_{\rm tot} =  (1-3J) N_{\triangle},
\end{align}
which is the minimum of energy determined in Eq.~(\ref{eq:minNRJ}).
This completes the proof that no other phases exist for $J>1/3$.

\section{Details of global updates}
In this section, we will present the details of three types of global update algorithms. 
We first explain the loop update~\cite{barkema1998,moller2009,chern2012}, which accelerates the relaxation of the system 
in low-temperature region for $-1<J<0$.  
This algorithm is useful when the magnetic charges have a long-range order, while the spins are left disordered.
In this case, local spin flip creates a high-energy state by changing the charge state, and is hardly accepted, accordingly.
Meanwhile, the loop update relaxes the spin configuration effectively, without changing 
charge configuration.
Secondly, we explain how to get rid of metastable defects 
by using the string update which causes the pair annihilation of the defects.  
Finally, we introduce a new worm algorithm, which is implemented in low-temperature region for $0<J<\frac{1}{3}$. 
This algorithm transports a charge between different clusters, and relaxes
both spin and charge configurations without changing the energy. 

\subsection{Loop update}
\label{sec:loopupdate}
\begin{figure}[!htb]
\begin{center}
\includegraphics[bb=0 0 703 261,width=\linewidth]{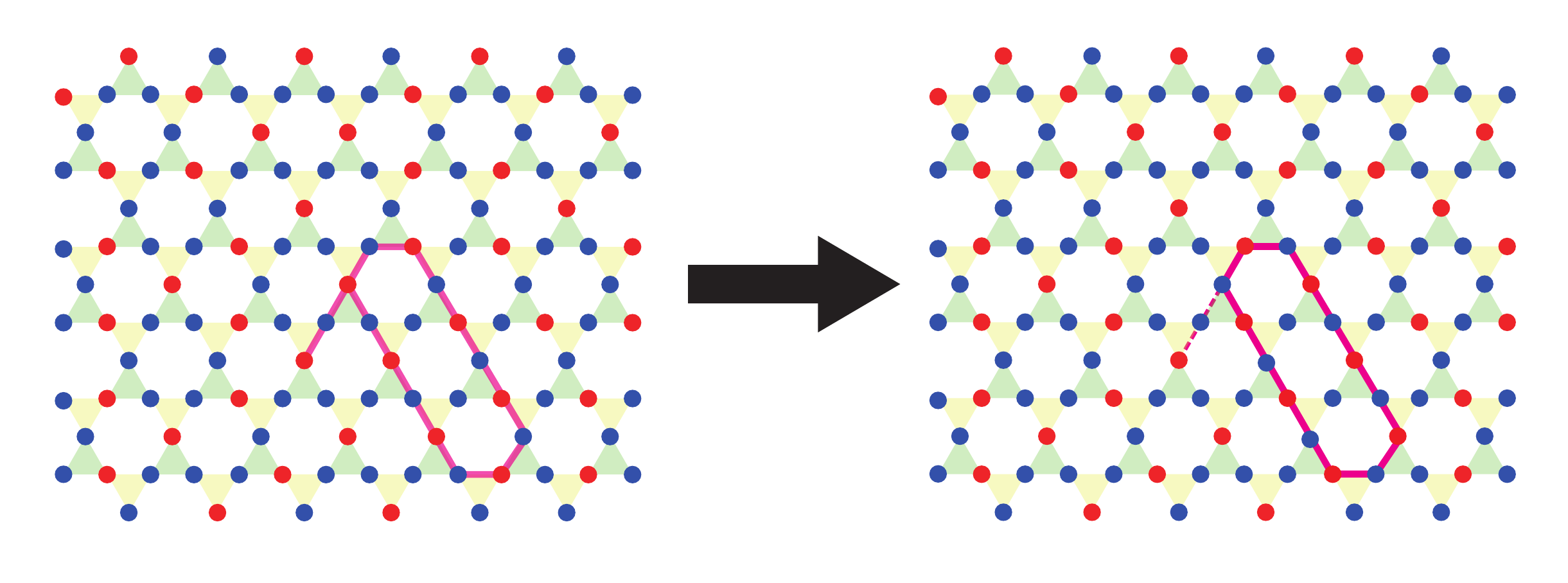}
\vspace{-10pt}
\caption{Schematic picture of the loop update, formed by a closed loop and a dangling tail.} 
\label{fig:loop}
\end{center}
\end{figure}

In the CSL state with staggered charge order, which is stabilized for $-1<J<0$, 
local spin updates that change the charge configuration are hardly accepted in the low-temperature regions,
since such updates cost a lot of energy. Hence, we need an algorithm that relaxes spin configuration without changing the charge configuration. 
The loop update serves this purpose~\cite{barkema1998,moller2009,chern2012}. This algorithm is illustrated in Fig.~\ref{fig:loop} and consists of finding a closed loop of spins aligned alternately, like $\uparrow \downarrow \uparrow \downarrow \cdots$
During the construction of the loop, because of the underlying \textit{staggered} charge order, 
the spins in the loop are successively chosen with probability $1$ (no choice) and 0.5 (a choice between two spins). Once a given spin is visited twice, the process stops and the global path of spins can be divided into a closed loop and a dangling tail. The spins in the closed loop are flipped, while the ones in the dangling tail are left unchanged. Since the charge configuration is left unchanged, there is no energy cost and this loop update is always accepted in Monte Carlo simulations.
%---------%
%---------%

\subsection{String update}
\label{sec:stringupdate}
\begin{figure}[!htb]
\begin{center}
\includegraphics[bb = 0 0 700 240,width=\linewidth]{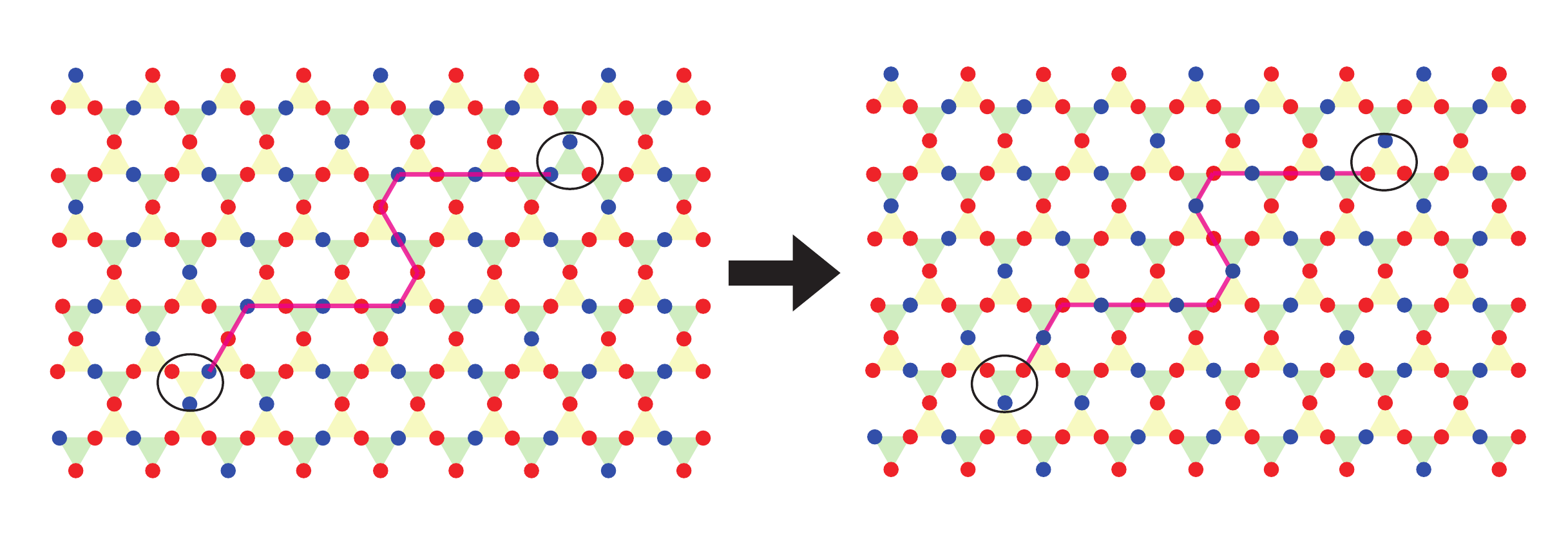}
\vspace{-10pt}
\caption{Schematic picture of the string update.} 
\label{fig:string}
\end{center}
\end{figure}

In the low temperature region of $-1<J<0$, where the staggered charge order is stabilized, same-charge clusters are
sometimes left unrelaxed, and remain as metastable defects 
[see the black circle in Fig. \ref{fig:string}].
These defects are difficult to remove by single spin flip,
since local spin updates conserve charge locally. 
This would result in triple charges which are unfavored for $-1<J<0$.
To eliminate these defects, we need to pair-annihilate two defects of opposite sign, which exist in spatially separate places. 

To this aim, we implement a string update algorithm. 
In this update, 
we first choose a triangle randomly and select one of the majority spins as a starting site.
Then, we make a string on which spins are aligned alternately: 
$\uparrow \downarrow \uparrow \downarrow \cdots $, 
in the same way as in the loop update algorithm introduced in Sec.~\ref{sec:loopupdate}.
However, in the string update, we do not make a closed loop, but terminate a string with open ends, at randomly-chosen length. If the string update is accepted by a Metropolis argument, then all the spins along that string are flipped. 
This attempt transports charge from one end to the other. 
Accordingly, if the chosen string has defects of opposite signs on both ends, they can be pair-annihilated [see Fig.~\ref{fig:string}].
To eliminate defects, however, we need many attempts, since the starting site and the length of the string are chosen randomly.

%-------------------------------------------------------------------%
%--------------------------------------------------------------------%

\subsection{Worm algorithm}
In the hexamer CSL realized for
$0<J<\frac{1}{3}$, low-temperature configurations are locally constrained by the attraction between same-sign charges. 
Single-spin-flip dynamics do not suffice to explore the extensively degenerate ensemble of ground states. 
This ergodicity loss is further complicated by the absence of any long-range charge order, which prevents the use of traditional loop algorithms~\cite{barkema1998} as described in section~\ref{sec:loopupdate}. 
To restore ergodicity, a new kind of worm algorithm is thus needed, able to update both spin and charge configurations. 

\begin{figure}[!htb]
\begin{center}
\includegraphics[bb= 0 0 699 257,width=\linewidth]{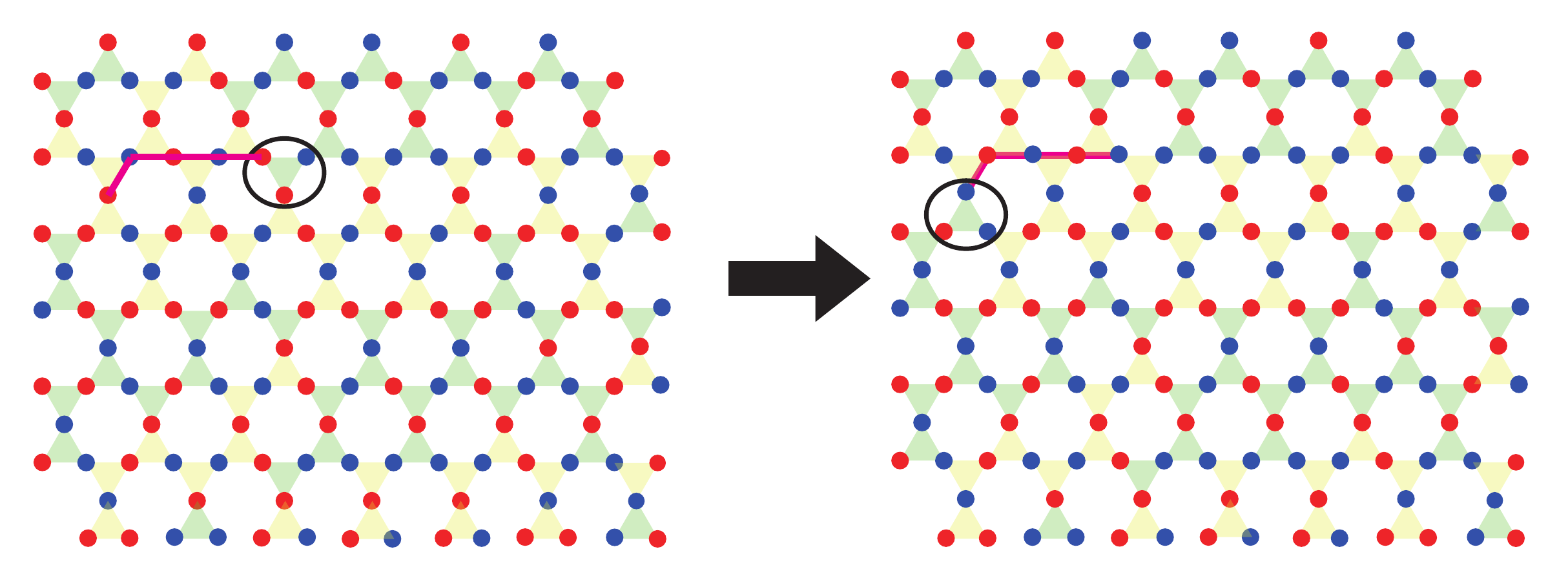}
\vspace{-10pt}
\caption{Schematic picture of the worm algorithm.} 
\label{fig:worm}
\end{center}
\end{figure}

To introduce the main idea of this algorithm, a careful understanding of the structure of same-charge clusters is required.
Triangles within a cluster can be classified into three groups, according to how many like-charges they are neighboring:
\begin{itemize}
\item ``end-point'' triangles, sitting at the extremities of a cluster, with only one neighboring triangle carrying the same topological charge;
\item ``middle'' triangles, forming the core of the branches of the clusters, with two neighboring same-sign triangles;
\item ``branching triangles'', surrounded by three neighboring same-sign triangles.
\end{itemize}
The main idea of the algorithm developed here is to detach an end-point charge from a given cluster, and to stick it back either somewhere else onto the same cluster, or to a different cluster of the same sign. Along the path of this end-point charge, the intervening cluster of opposite sign is also updated. This is done by an extension of
the string update algorithm introduced in section \ref{sec:stringupdate}.
As illustrated in Fig.~\ref{fig:worm}, we firstly choose an end-point triangle of positive (negative) sign randomly. This triangle is by definition surrounded by two triangles of opposite sign; the string starts from one of the two corresponding boundary spins with equal probability.
However, in contrast to the string update introduced in Sec. \ref{sec:stringupdate}, the string propagates within the neighboring cluster of opposite sign (and not on a background of staggered charge order), and is not terminated at random. Alternatively, we monitor the type of triangles at each step, and stop the update as soon as we reach a middle triangle of negative (positive) sign. At this point, we flip the constructed open string of spins.
As a result, a positive (negative) end-point charge can be moved around in the system [Fig. \ref{fig:worm}].

At the end of the process, the total energy of the system is preserved, such that the attempt is always accepted by the Metropolis rule.
Furthermore, the key in this algorithm lies in the necessary respect of detailed balance.
Indeed, while the initial triangle always change from an ``end-point'' to a ``middle'' state, the final triangle always follows the reverse change of status. 
And since we stop the string update at the {\it first} middle triangle encountered, 
it guarantees the necessary one-to-one correspondence between the string update and its reversed process. 
These conditions ensure the detailed balance of stochastic processes.

%---------%
%---------%

%---------%
\begin{figure}[!htb]
\begin{center}
\includegraphics[bb = 0 0 609 450,width=\linewidth]{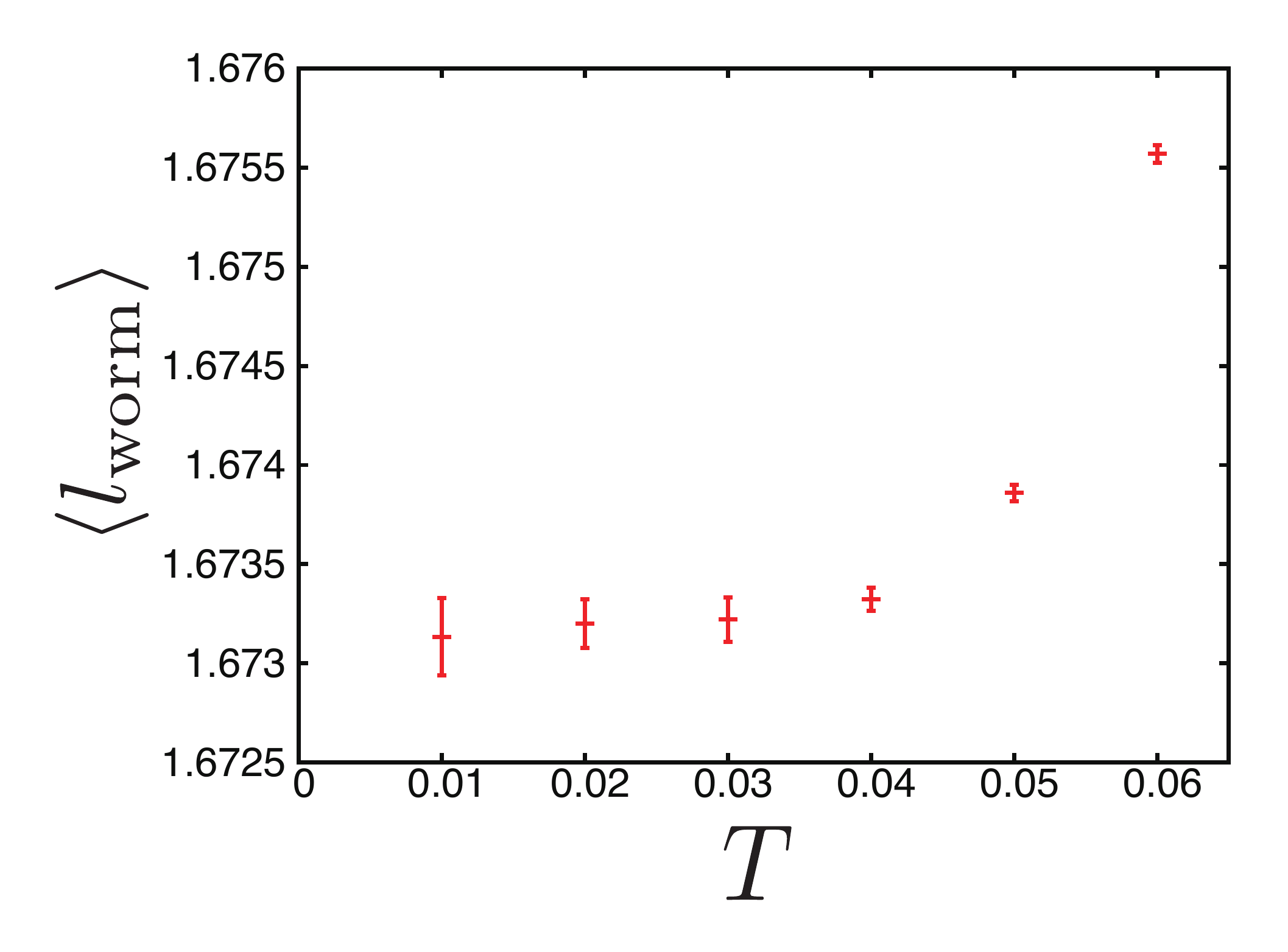}
\vspace{-10pt}
\caption{Temperature dependence of the average length of worms.} 
\label{fig:worml}
\end{center}
\end{figure}
%---------%
Below, we show the benchmark of the algorithm.
The average length of worms, $\langle \ell_{\mathrm{worm}} \rangle$, is shown in Fig. \ref{fig:worml}.
One can see that the average length is almost temperature-independent and its value is about $1.67$,
which indicates that the worm algorithm changes the states locally rather than globally.

\begin{figure}[!htb]
\begin{center}
\includegraphics[bb= 0 0 550 223,width=\linewidth]{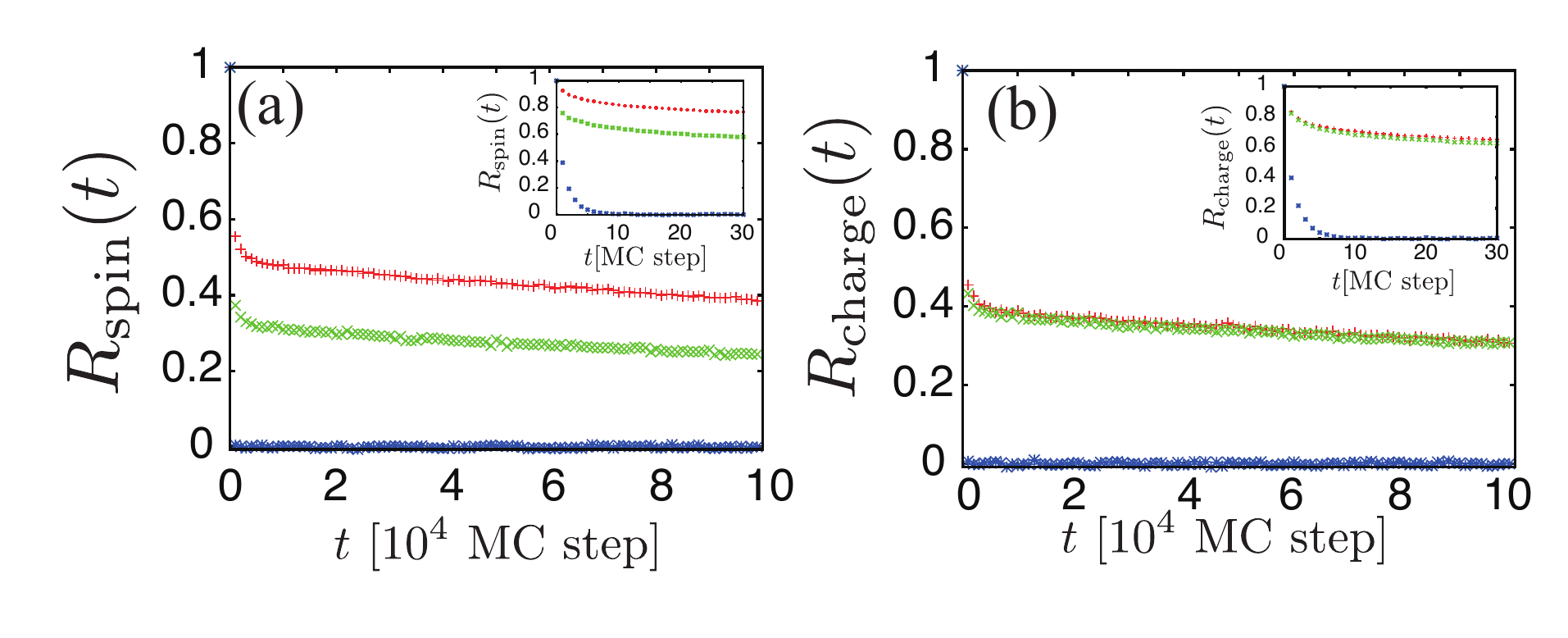}
\vspace{-10pt}
\caption{The autocorrelation of the spin and charges. $(N_{\mathrm{single}},N_{\mathrm{loop}},N_{\mathrm{worm}})$ for red, green, and blue dots 
are described in the text. } 
\label{fig:ac}
\end{center}
\end{figure}

Nevertheless, the worm algorithm improves the relaxation drastically;
to show its quantitative evidence, we calculate the autocorrelation.   
The autocorrelations of spins, $R_{\spin} \equiv \frac{\sum_i [\sigma_i^z (0) \sigma_i^z (t) ]}{N_{\mathrm{site}} }$, and charges, $R_{\mathrm{charge} } \equiv \frac{ \sum_{p \in \bigtriangleup} [Q_p(0) Q_p(t) ]}{N_{\bigtriangleup} } $, 
are shown in Fig. \ref{fig:ac}, respectively.
Here one Monte Carlo step consists of $N_{\mathrm{single}}$ single spin flip, $N_{\mathrm{loop} }$ loop updates, and $N_{\mathrm{worm} }$ worm updates.
In Fig. \ref{fig:ac}, 
these numbers are chosen as 
$(N_{\mathrm{single}},N_{\mathrm{loop}},N_{\mathrm{worm}}) = (N_{\mathrm{site}},0,0)$ for red dots, 
$(N_{\mathrm{single}},N_{\mathrm{loop}},N_{\mathrm{worm}}) = (N_{\mathrm{site}},N_{\mathrm{site}}/8,0)$ for green dots,
and $(N_{\mathrm{single}},N_{\mathrm{loop}},N_{\mathrm{worm}}) = (N_{\mathrm{site}},0,N_{\mathrm{site}})$ for blue dots,
respectively.
One can clearly see that $R_{\spin}$ and $R_{\charge}$ 
approach to zero very rapidly, if we introduce the new worm algorithm. 
On the other hand, the loop update algorithm is not efficient in hexamer CSL state. 
In particular, $R_{\charge}$ is unchanged by loop update, 
since it does not change the charge configuration.

\begin{figure}[!htb]
\begin{center}
\includegraphics[bb= 0 0 835 333,width=\linewidth]{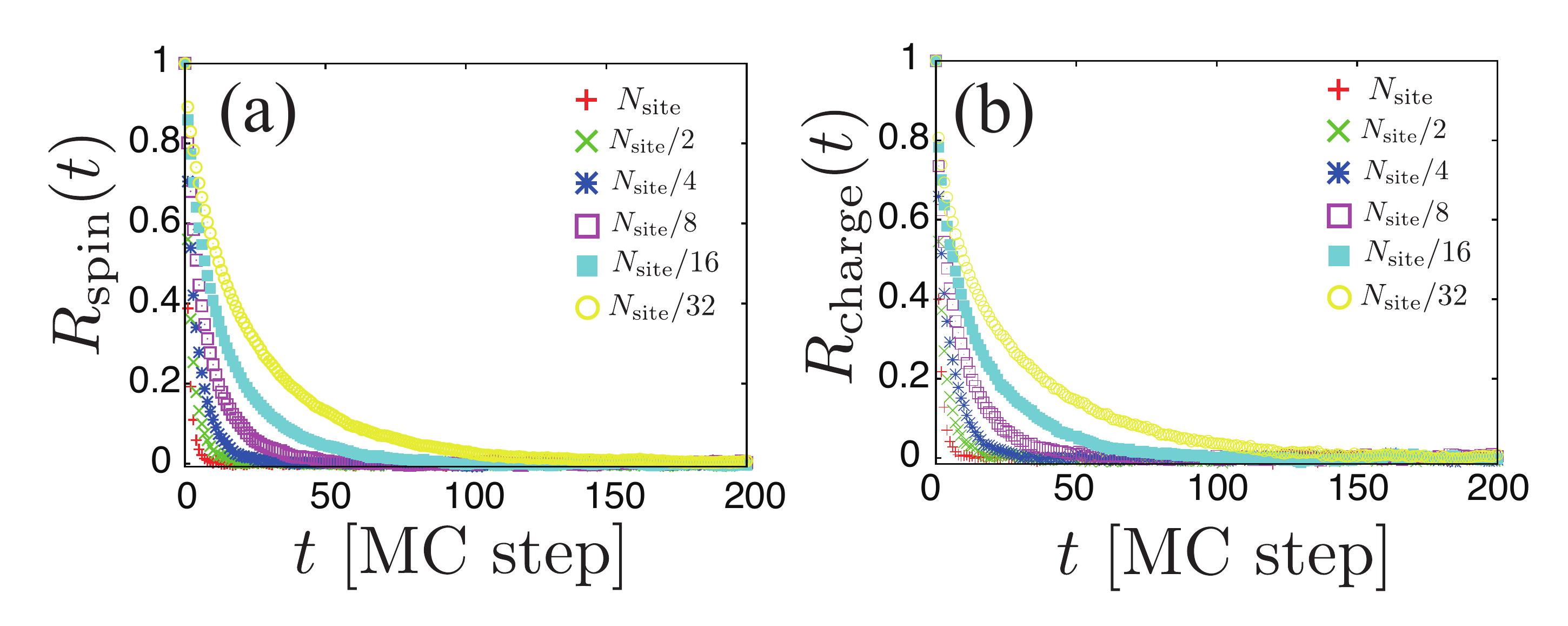}
\vspace{-10pt}
\caption{The autocorrelation of the spin and charges for various values of $N_{\mathrm{worm}}$.} 
\label{fig:ac2}
\end{center}
\end{figure}

$N_{\mathrm{worm}}$ dependence of the autocorrelations are shown in Fig. \ref{fig:ac2}.
Clearly, the larger $N_{\mathrm{worm}}$ becomes, the faster the relaxation is achieved.
Even in the smallest number of $N_{\mathrm{worm}}(=\frac{N_{\mathrm{site}}}{32})$,
$R_{\spin}$ and $R_{\charge}$ go to zero within 150 Monte Carlo steps. 
This demonstrates that the worm algorithm is indeed a powerful technique to relax the system into equilibrium. 

%---------%
%---------%
%---------%
%---------%

%---------%
\section{Pauling's argument in new classical spin liquid}
%---------%
\begin{figure}[!htb]
\begin{center}
\includegraphics[bb= 0 0 263 233,width=\linewidth]{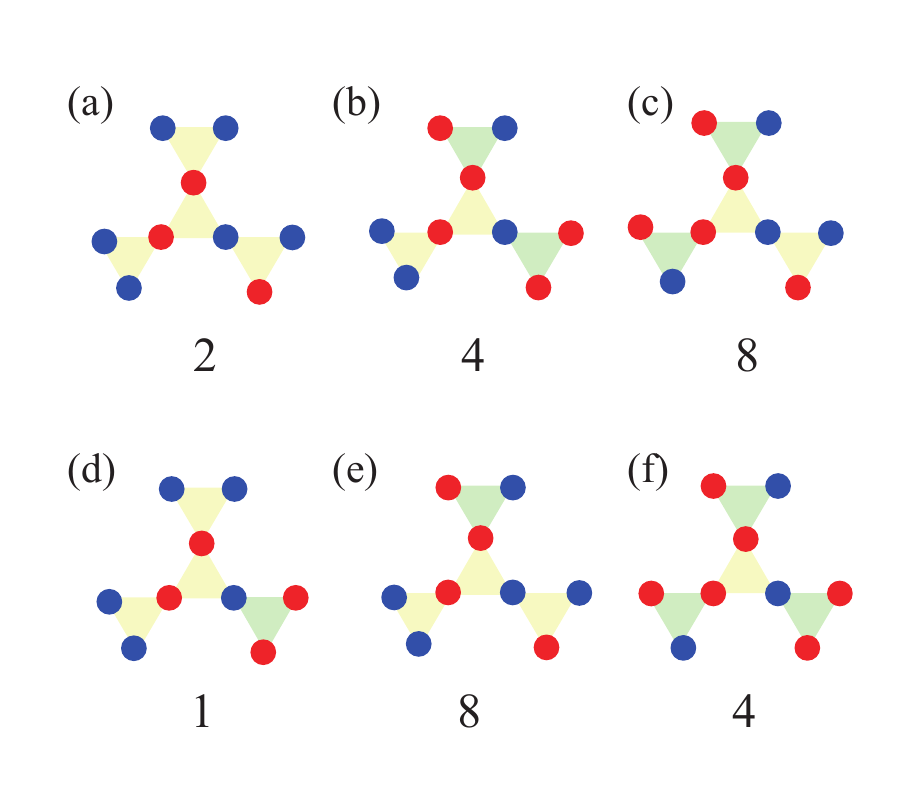}
\vspace{-10pt}
\caption{Charge configurations on a four-triangle cluster.
Here we fix the spin configuration on the central triangle. 
Spin degeneracy of each configuration is shown below each snapshot. }  
\label{fig:pauling}
\end{center}
\end{figure}
%---------%
In this section, we discuss the residual entropy of hexamer CSL
on the basis of Pauling's argument. 
As we mentioned in the main text, 
the idea is that we consider a four-triangle cluster consisting of nine spins 
as a unit [see Fig. \ref{fig:pauling}]. Suppose that the central triangle has a charge $Q=+1$, without any loss of generality. For a given fixed spin configuration on the central triangle, charge configurations of the other three triangles are classified in six categories, as shown in Figs. \ref{fig:pauling}(a)-(f). 
Since we are in the hexamer CSL phase, there are no triple charges. The six types of  charge configurations are classified as follows:
\begin{enumerate}
\item[(a)] Three same-sign-charge pairs. 
\item[(b)] One same-sign-charge pair and two opposite-sign-charge pairs: The minority spin of the central triangle connects opposite-sign charges.
\item[(c)] One same-sign-charge pair and two opposite-sign-charge pairs: The minority spin of the central triangle connects same-sign charges. 
\item[(d)] Two same-sign-charge pairs and one opposite-sign-charge pair: The minority spin of the central triangle connects opposite-sign charges.
\item[(e)] Two same-sign-charge pair and one opposite-sign-charge pair: The minority spin of the central triangle connects same-sign charges.
\item[(f)] Three opposite-sign-charge pairs.  
\end{enumerate}
Each charge configuration has a spin degeneracy (with respect to the six exterior spins), which is also presented in the figures. 

Among these configurations, (d) and (e) have two like-charge pairs, 
which is equal to the average number of like-charge pairs in hexamer CSL (see the main text). 
Thus, the residual entropy can be estimated
by considering the probability that each four-triangle cluster 
has a configuration of either (d) or (e). 
Note that we also have to consider the case that the central triangle has $Q=-1$. 
That probability, $P_{\mathrm{d}+\mathrm{e}}$, can be derived as follows:
\begin{enumerate}
\item A four-triangle cluster consist of nine spins, so there are $2^9$ states.
\item A central triangle has either $Q=+1$ and $Q=-1$, and the number of such spin configuration is six.
\item Once we fix the spin configuration on a central triangle, the number of states with 
patterns of (d) and (e) is $1+8=9$.
\end {enumerate}
Then we obtain
\begin{equation}
P_{\mathrm{d}+\mathrm{e}} = \frac{6 \times 9}{2^9} = \frac{3^3 }{2^8}.
\end{equation}
The corresponding residual entropy becomes
\begin{equation}
S^{\mathrm{HCSL}}_P = \frac{1}{N_{\mathrm{site}}} \log \left( 2^{N_{\mathrm{site}}} P_{\mathrm{d}+\mathrm{e}}^{N_c}  \right) = \frac{1}{6} \log \left(\frac{3^3}{2^2} \right) \sim 0.318, 
\end{equation}
where $N_c$ denotes the total number of four-triangle clusters, 
\begin{equation}
N_c = \frac{1}{4} N_{\triangle} = \frac{N_{\mathrm{site}}}{6},
\end{equation}
and $2^{N_{\mathrm{site}}}$ is the total number of states.
The obtained value of the residual entropy is very close to the one obtained by the Monte Carlo simulation. 

%%%%%%%%%%%%%%%%%%%%%